\documentclass[fleqn,usenatbib]{mnras}
\usepackage{newtxtext,newtxmath}
\usepackage[T1]{fontenc}

\usepackage{ae,aecompl}
\usepackage{comment}

\usepackage{microtype}
\usepackage{amsmath}
\usepackage{graphicx}
\usepackage{lscape}
\usepackage{color}
\usepackage[normalem]{ulem}
\graphicspath{{pics/}}
\newcommand {\bc}{\begin {center}}
\newcommand {\ec}{\end {center}}
\newcommand {\be}{\begin {equation}}
\newcommand {\ee}{\end {equation}}
\newcommand {\beq}{\begin {eqnarray}}
\newcommand {\eeq}{\end {eqnarray}}
\newcommand {\ergs}{{\rm erg\,s$^{-1}$}}

\usepackage{caption}
\usepackage{subcaption}

\definecolor{mypink1}{rgb}{0.858, 0.188, 0.478}

\newcommand{\flux}{erg\,s$^{-1}$\,cm$^{-2}$}

\newcommand{\rx}{{RX\,J0440.9$+$4431}}

\title[Supercritical state of \rx]
{\rx: another supercritical X-ray pulsar}

\author[A.~Salganik et al.]
{Alexander~Salganik,$^{1,2}$\thanks{E-mail: alsalganik@gmail.com} 
Sergey~S.~Tsygankov,$^{3}$ Victor Doroshenko,$^{4}$ 
Sergey~V.~Molkov,$^{2}$ 
\newauthor
Alexander A. Lutovinov,$^{2}$ 
Alexander~A.~Mushtukov,$^{5}$
and Juri~Poutanen$^{3}$
\\
$^1$Department of Astronomy, Saint Petersburg State University, Saint-Petersburg 198504, Russia\\
$^2$Space Research Institute of the Russian Academy of Sciences, Profsoyuznaya Str. 84/32, Moscow 117997, Russia\\
$^3$Department of Physics and Astronomy,  FI-20014 University of Turku, Finland\\
$^4$Institut für Astronomie und Astrophysik, Universit\"at T\"ubingen, Sand 1, D-72076 T\"ubingen, Germany\\
$^5$Astrophysics, Department of Physics, University of Oxford, Denys Wilkinson Building, Keble Road, Oxford OX1 3RH, UK 
}
\pubyear{2023}

\begin{document}
\label{firstpage}
\pagerange{\pageref{firstpage}--\pageref{lastpage}}
\maketitle

%%%%%%%%%%%%%%%%%%%%%%%%%%%%%%%%%%%%%%%%%%%%%%%%%%%%%%%%%%%%%%%%%%%%%%%%%%%%%%
%% Abstract, Keywords and contact details                                   %%
%%%%%%%%%%%%%%%%%%%%%%%%%%%%%%%%%%%%%%%%%%%%%%%%%%%%%%%%%%%%%%%%%%%%%%%%%%%%%%
\begin{abstract}
In the beginning of 2023 the Be transient X-ray pulsar \rx\ underwent a fist-ever giant outburst observed from the source peaking in the beginning of February and reaching peak luminosity of $\sim 4.3\times10^{37}$~\ergs.  Here we present the results of a detailed spectral and temporal study of the source based on \textit{NuSTAR}, \textit{INTEGRAL} and \textit{NICER} observations performed during this period and covering wide range of energies and luminosities. 
We find that both the pulse profile shape and spectral hardness change abruptly around $\sim2.8\times10^{37}$~\ergs, which we associate with a transition to super-critical accretion regime and erection of the accretion column. 
The observed pulsed fraction decreases gradually with energy up to 20\,keV (with a local minimum around fluorescence iron line), which is unusual for an X-ray pulsar, and then rises rapidly at higher energies with the pulsations significantly detected up to $\sim120$~keV. The broadband energy spectra of \rx\ at different luminosity states can be approximated with a two-hump model with peaks at energies of about 10--20 and 50--70\,keV previously suggested for other pulsars without additional features. In particular an absorption feature around 30~keV previously reported and interpreted as a cyclotron line in the literature appears to be absent when using this model, so the question regarding the magnetic field strength of the neutron star remains open. Instead, we attempted to estimate field using several indirect methods and conclude that all of them point to a relatively strong field of around $B\sim 10^{13}$~G.
\end{abstract}


%%%%%%%%%%%%%%%%%%%%%%%%%%%%%%%%%%%%%%%%%%%%%%%%%%%%%%%%%%%%%%%%%%%%%%%%%%%%%%
%% Introduction                                                             %%
%%%%%%%%%%%%%%%%%%%%%%%%%%%%%%%%%%%%%%%%%%%%%%%%%%%%%%%%%%%%%%%%%%%%%%%%%%%%%%
\section{Introduction}
\label{sec:intro}
\rx\ was discovered by \citet{Motch1997} as part of the ROSAT galactic plane survey and classified as a Be/X-ray binary based on properties of the identified optical counterpart BSD 24-491/LS V +44 17, which has spectral type B0.2Ve \citep{Reig2005b}. 
\citet{Reig1999} showed that the compact object in the system is an X-ray pulsar (see \citealt{Mushtukov22}, for a recent review), with the pulse period of $P_{\rm spin}$ = $202.5\pm0.5$~s. No previous outburst activity in combination with a low luminosity of $\sim10^{34-35}$ \ergs\ and a large pulse period led \citet{Reig1999} to suggest that \rx\ is a member of a not so numerous family of persistent low-luminosity Be/X-ray binaries. 

In March 2010 the MAXI all-sky monitor detected, however, the first outburst from the source \citep{Morii2010} which peaked at $3.9\times10^{36}$\,\ergs\ (3--30\,keV; \citealt{Usui2012}) suggesting that source is actually a transient with relatively high quiescent luminosity $L_x\sim10^{34}$ \ergs. Assuming the distance to \rx\ of $2.44^{+0.06}_{-0.08}$~kpc \citep{Bailer2021}, the observed luminosity remained so far below $10^{37}$\,\ergs\ typical for the Type I outbursts of Be X-ray binaries \citep{Morii2010, Tsygankov2011, Krivonos2010b, Finger2010,Reig2011}. 

\rx\ soft X-ray spectra measured with different X-ray observatories can be approximated with a combination of power-law and black-body components with the inclusion of the 6.4\,keV iron line \citep{Usui2012, Tsygankov2012, Palombara2012}. At higher energies an absorption feature at $\sim$30\,keV interpreted as a possible cyclotron line was reported \citep{Tsygankov2012}, which implies magnetic field of $3\times10^{12}$~G. This estimate was also consistent with the magnetosphere size estimated based on the frequency of a break in the source power density spectra \citep{Tsygankov2012}. However, subsequent research has raised doubts about the existence of this spectral feature \citep{Ferrigno2013}. 

The pulse profile of \rx\ has been reported to have a simple sine-like shape \citep{Reig1999} modified at soft energies by a narrow dip, which can be interpreted as absorption by the accretion stream or the accretion column \citep{Usui2012, Tsygankov2012}. The orbital parameters of the system are not known. However, based on \textit{Swift}/BAT telescope data, the presence of an orbital variations with $\sim$155\,d period was reported by \citet{Tsygankov2011}, and the period of $150.0\pm0.2$\,d was measured more precisely by \citet{Ferrigno2013}.

%======================================================
\begin{figure*}
    \centering
    \includegraphics[width=1.99\columnwidth]{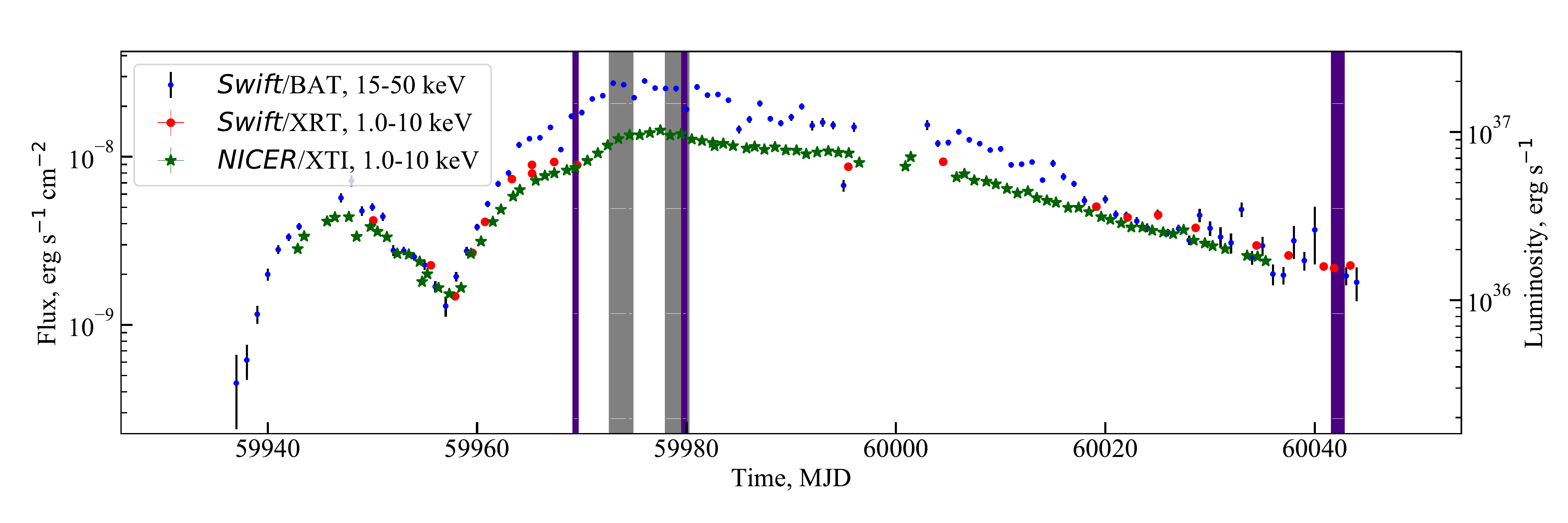}
 	\caption{Light curve of \rx\ based on the NICER/XTI, {\it Swift}/XRT and {\it Swift}/BAT data. Vertical strips represent intervals of the \textit{NuSTAR} (violet colour) and \textit{INTEGRAL} (grey colour) observations. The luminosity was calculated assuming distance to the source of 2.44 kpc. To convert the BAT light curve into flux, the Crab count rate of 0.22 count~cm$^{-2}$~s$^{-1}$ was assumed \citep{Romano2014}.}
	\label{fig:lightcurve}
\end{figure*}
%======================================================

The most recent Type I outburst of \rx\ was detected on 2022 December 29 by the MAXI all-sky monitor \citep{Nakajima2022}. Based on the \textit{Swift}/BAT data, the outburst peaked on 2023 January 4 with the flux of $\sim 0.6$ Crab in the 15--50\,keV band ($L\sim5\times10^{36}$ \ergs). After that, the source began to fade rapidly, until a trend reversed around January 13 with an onset a giant (Type II) outburst eventually peaking at around $\sim2.2$ Crab in the 15–50 keV band ($L\sim2\times10^{37}$\ergs) on 2023 Jan 20 \citep{Pal2023}. 
Several follow-up observations by \textit{NuSTAR} and \textit{INTEGRAL} were conducted on MJD 59969 \citep{Salganik2023}, 59979, 60041, and MJD 59972, 59977, respectively.
Here, we report the results of analysis of this data and results of a monitoring campaign with \textit{Swift} and NICER. 

\begin{table}
\caption{Summary of the broadband observations.}
\begin{tabular}{lccc}
\hline
Observation & MJD   & Exposure, ks & Observatory       \\
\hline
NuObs1 & 59969 & 11.7         & \textit{NuSTAR}   \\
IntObs1        & 59972 & 99.7         & \textit{INTEGRAL} \\
IntObs2        & 59977 & 115.2        & \textit{INTEGRAL} \\
NuObs2 & 59979 & 8.8          & \textit{NuSTAR}   \\
NuObs3 & 60041 & 47.4         & \textit{NuSTAR}  \\
\hline
\end{tabular}
\label{table:observations}
\end{table}

\section{Data analysis}
\label{sec:data}
In our work, we utilized data from the NICER, \textit{Swift}, \textit{NuSTAR}, and \textit{INTEGRAL} observatories to study \rx\ temporal and spectral properties throughout the outburst. The NICER and \textit{Swift} data allow to study evolution of spectra and pulse profiles of the source in soft X-ray band in wide range of luminosity whereas the broadband \textit{INTEGRAL} and \textit{NuSTAR} observations allowed us to study \rx\ in wide range of energies. A summary of the broadband observations is presented in Table~\ref{table:observations}.

\subsection{\textit{NuSTAR} observatory}
\textit{NuSTAR} consists of two identical co-aligned X-ray focal plane modules (FPM) called FPMA and FPMB \citep{Harrison2013}.
\textit{NuSTAR} observatory provides wide band 3--79~keV data with a high sensitivity in the hard X-ray energy band, a good energy resolution of 400\,eV at 10\,keV, and angular resolution of 18 arcsec (FWHM).  \textit{NuSTAR} observations of \rx\ were performed on 2023 January 25 (ObsID 90901302002; MJD 59969),  February 4  (ObsID 90901304002; MJD 59979), and April 7 (ObsID 90901313002; MJD 60041) during the rising, peak, and decline parts of the outburst respectively (see Table~\ref{table:observations}). From now on, we will refer to them in the text simply as NuObs1, NuObs2 and NuObs3. To extract events, circular regions with radii of 60, 65 and 40 arcsec were used for the source region and 120 arcsec for the background region to maximize signal-to-noise ratio at high energies for NuObs1, NuObs2 and NuObs3, respectively.

\textit{NuSTAR} observations were processed in accordance with the official guidelines.\footnote{\url{https://heasarc.gsfc.nasa.gov/docs/nustar/analysis/nustar_swguide.pdf}}
The package {\sc heasoft} version 6.31.1 and calibration files {CALDB version 20211202 were used for processing. The spectra and light curves were extracted using the \texttt{nuproducts} procedure, which is part of the \texttt{nustardas} pipeline. 
The spectra after background subtraction were rebinned to have at least 25 counts per energy bin using \texttt{grppha} utility.
The light curves of FPMA and FPMB were barycentric corrected using the \texttt{barycorr} procedure and after background subtraction were co-added using the \texttt{lcmath} task in order to improve the statistics. No corrections were applied to account for binary motion as the orbit of the source is not known at the time of writing.  

\subsection{{\it Swift} observatory}

To study the evolution of flux (see Fig.~\ref{fig:lightcurve}) and spectrum throughout the outburst, a monitoring campaign with the XRT telescope \citep{Burrows2005} onboard the \textit{Neil Gehrels Swift} Observatory \citep{Gehrels2004}  consisting of 16 observations (ObsIDs 00089583001--5, 7--10, 19, 22--31, 01150107000) was triggered. The observations cover time interval MJD 59950--60043 and were performed in the PC and WT modes depending on the count-rate in a given observation. The observations potentially affected by the pile-up were excluded from the analysis. Data analysis software\footnote{\url{https://www.swift.ac.uk/user_objects/}} \citep{Evans2009} provided by the UK Swift Science Data Center was used to extract the source spectrum in each separate observation. Again, \textit{Swift}/XRT spectra were rebinned to have at least 25 counts per energy bin. 
% We fixed the hydrogen column density $N_{\rm H}$  at ??? (the best-fit value obtained from the broadband spectrum, see Sect.~\ref{sec:spectrum}) for all \textit{Swift}/XRT observations. 

\subsection{\textit{INTEGRAL} observatory}

\textit{INTEGRAL} \citep{2003A&A...411L...1W} performed two observations of \rx\ throughout revolutions 2600 (MJD 59972--59974) and 2602 (MJD 59977--59980), thereafter we will refer to them as IntObs1  and IntObs2.  In this work, we used data from the ISGRI detector \citep{2003A&A...411L.141L} of the IBIS telescope \citep{2003A&A...411L.131U} and the Joint European X-ray Monitor \citep[JEM-X; ][]{2003A&A...411L.231L} on board the observatory.

The INTEGRAL data were processed using the Off-line Scientific Analysis \citep[\textsc{OSA}, version 11.2;][]{2003A&A...411L..53C} software distributed by the \textit{INTEGRAL} Science Data Center (ISDC).  To reconstruct energy spectra for both JEM-X and IBIS telescopes we used the multi-messenger online data analysis platform \citep[MMODA;][]{2021A&A...651A..97N}. Before making the ISGRI light curves, we first recalculated the photon energies in the instrument's event files using up-to-date calibration data.  To do this, we launched the \textsc{OSA} procedure \texttt{ibis\_science\_analysis} to the COR level. After that, the arrival time of photons in resulted files were corrected to the barycenter of the Solar system. 
Next, using the \texttt{ii\_pif} procedure, we calculated the ``open part'' of each pixel (i.e. pixel-illumination-fraction, PIF) of the detector  for direction to \rx. Then, we constructed a light curve in a given energy band with a 1~s time resolution for ``fully open'' pixels and subtracted from it a light curve obtained from the ``fully closed'' pixels with the same time resolution, taking into account the number of pixels used for each light curve. In our case, when there is only one bright source in the field of view of the IBIS telescope and the hexagonal (HEX) pattern is used (that is, the source is always in the fully coded part of FoV), this is the simplest and most correct way to reconstruct a light curve.

\subsection{NICER observatory}

NICER is an externally attached payload on the International Space Station \citep{Gendreau2016}. NICER's main instrument is the X-ray Timing Instrument (XTI), operating in the 0.2--12\,keV energy range, with a large effective area of 1700\,cm$^{2}$ at 1\,keV, high sensitivity, and unprecedented time-tagging resolution of less than 300 ns. We used data from the monitoring campaign of \rx\ consisting of 88 observations covering MJD 59942-60036 (ObsIDs 5203610101-55,57-58; 5400690101-2, 6203610101-30).

NICER data were reduced using the {\sc nicerdas} software and {CALDB} version 20221001 according the official data analysis threads.\footnote{\url{https://heasarc.gsfc.nasa.gov/docs/nicer/analysis_threads/}} Barycentric correction was applied to the light curves using the \texttt{barycorr} procedure. All spectra were rebinned to have at least 25 counts per energy bin.
All data in this work were fitted in the XSPEC 12.12.1 \citep{Arnaud1996} using $\chi^2$ statistic. All errors are given at the 1$\sigma$ confidence level if not specified otherwise.

\section{Results}
Using the data described above we were able to study temporal and spectral properties of \rx\ at different time scales. Figure~\ref{fig:lightcurve} shows the light curve of the source based on the NICER/XTI, {\it Swift}/BAT\footnote{\url{https://swift.gsfc.nasa.gov/results/transients/weak/LSVp4417/}} and {\it Swift}/XRT observations. \textit{Swift}/XRT fluxes were estimated based on approximation of observed spectra using an absorbed power-law model whereas NICER spectra required  an addition of a blackbody component with temperature of $\sim2-3$ keV. 

\subsection{Timing analysis}

\begin{figure}
    \centering
    \includegraphics[width=1.0\columnwidth]{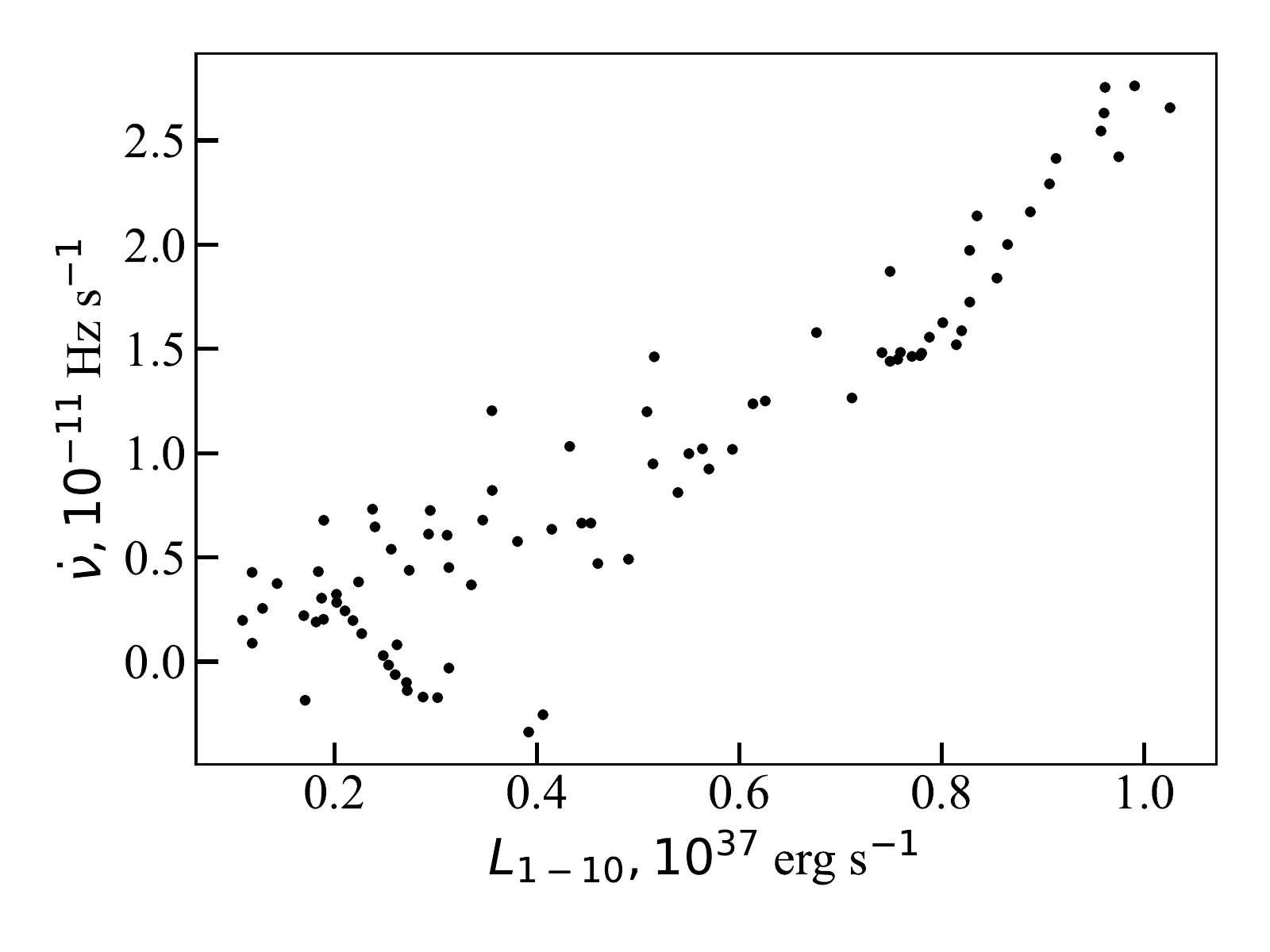}\\
    \caption{\rx\ spin-up rate $\dot{\nu}$ as a function of luminosity in the 1--10\,keV range  based on the NICER data. }
    \label{fig:spinup}
\end{figure}

\subsubsection{Pulse profile evolution with energy}
\label{sec:broad_timing}

To study the properties of \rx\ in a wide range of energies, we used data from the \textit{INTEGRAL} and \textit{NuSTAR} observatories. Because the mass accretion rate near the peak of the outburst was very high, the pulsar spin frequency changed significantly during the corresponding observations with the derivative $\dot{\nu}\sim(2-3)\times10^{-11}$ Hz s$^{-1}$, which corresponds to a period change of $\sim10^{-3}$ s for every 1000 s (see Fig.~\ref{fig:spinup}). To take this effect into account, we folded \textit{NuSTAR} and \textit{INTEGRAL} light curves taking into account period derivatives to produce the pulse profiles. The period and its derivative were obtained through cubic-spline-interpolation of periods from \textit{Fermi} GBM Accreting Pulsars Program (GAPP) \footnote{\url{https:// gammaray.nsstc.nasa.gov/gbm/science/pulsars/lightcurves/rxj0440.html}} \citep{Malacaria2020}.

\begin{figure*}
    \centering
    \includegraphics[width=0.68\columnwidth]{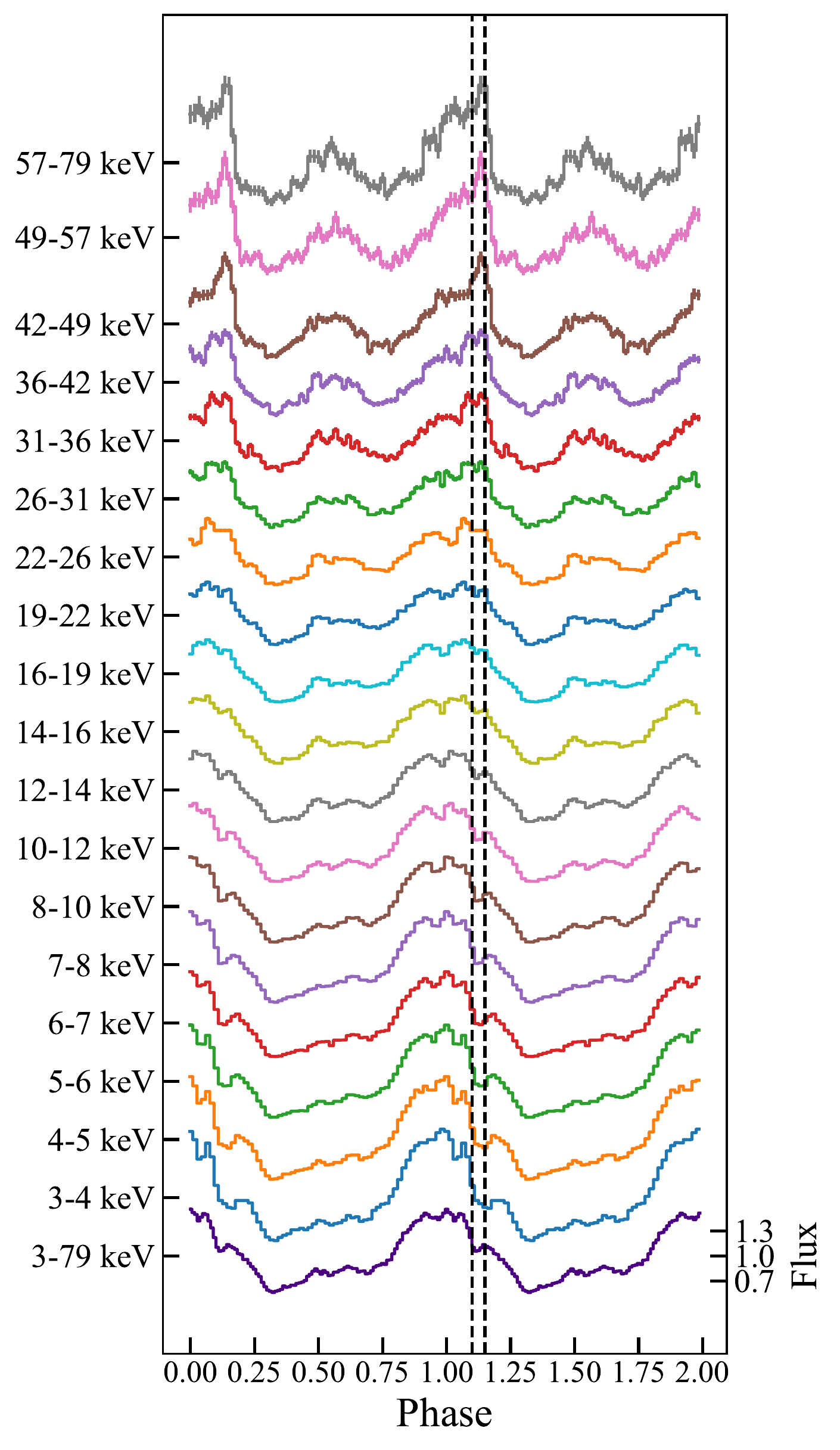}
    \includegraphics[width=0.68\columnwidth]{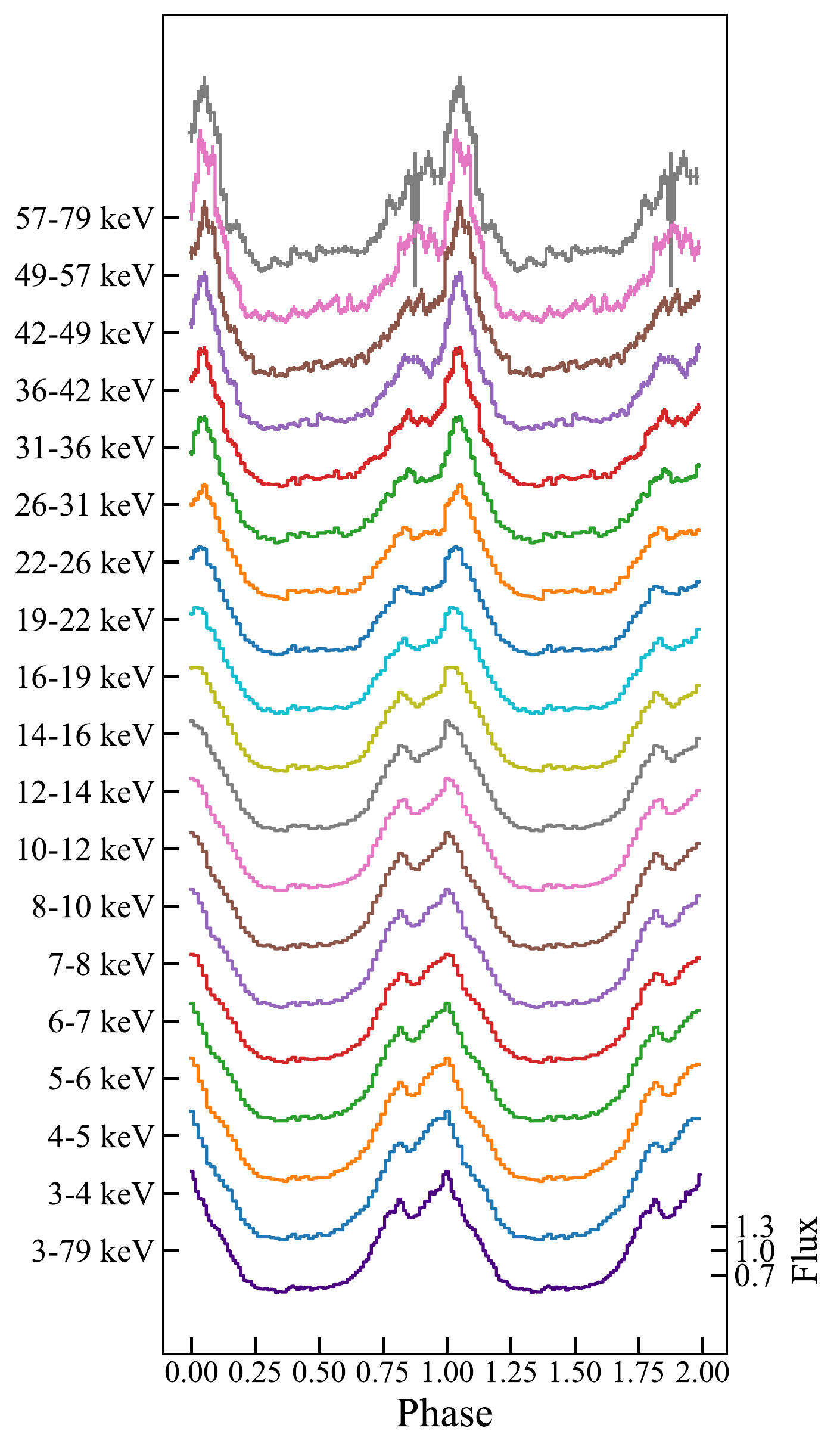}
    \includegraphics[width=0.68\columnwidth]{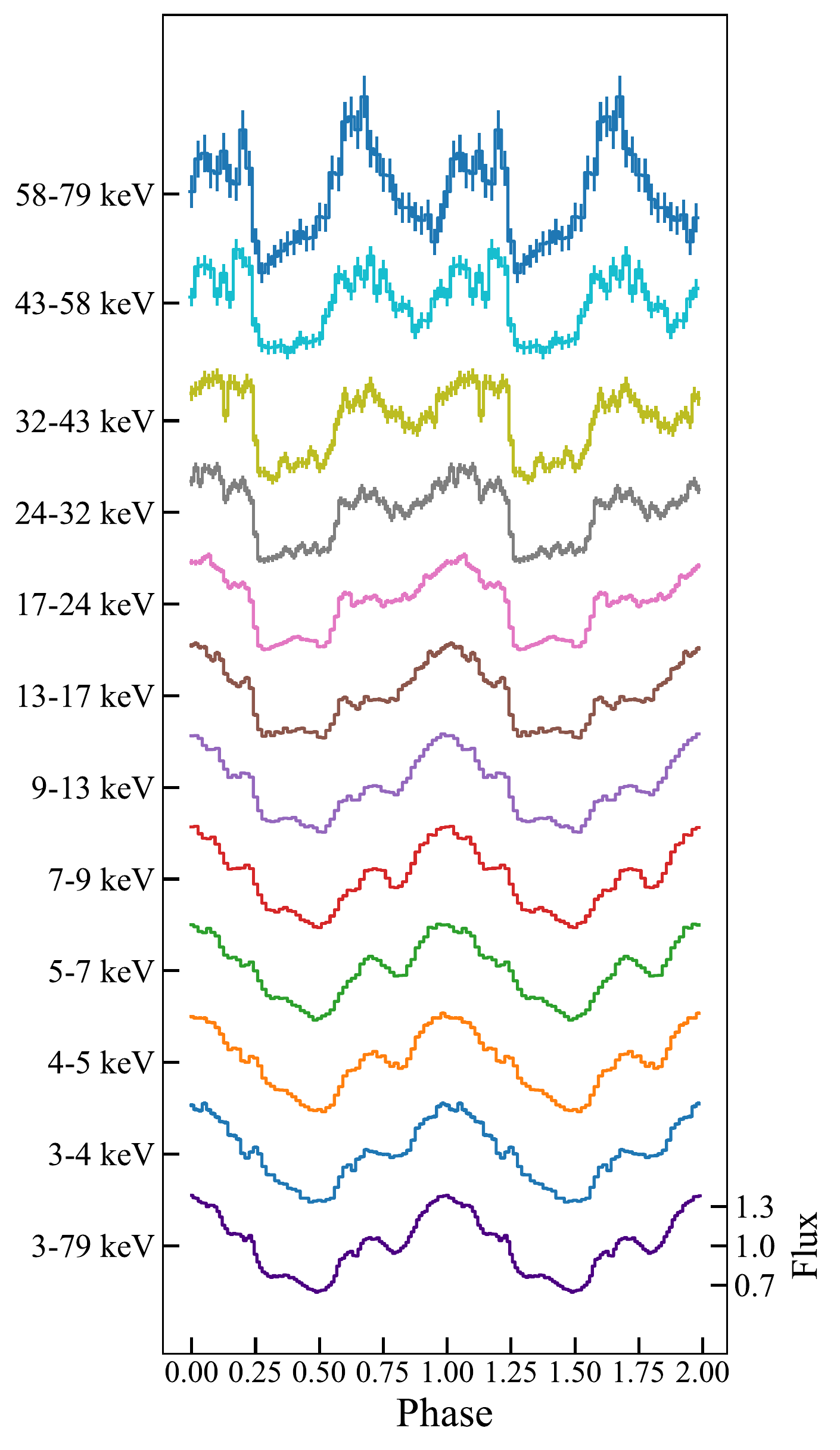}
\caption{Pulse profiles of \rx\ as a function of the energy based on the \textit{NuSTAR} data.   Two periods are shown.  Pulse profiles are spaced along the y-axis for better visualization convenience. Fluxes in each profile were normalized by their average. The phase of maximal flux was taken as the null phase for each observation. Black dashed lines represent the boundaries of the dip. 
Panels from left to right are for observations NuObs1, NuObs2 and NuObs3, respectively.  }
    \label{fig:profiles}
\end{figure*}

\begin{figure*}
    \centering
    \includegraphics[width=1.0\columnwidth]{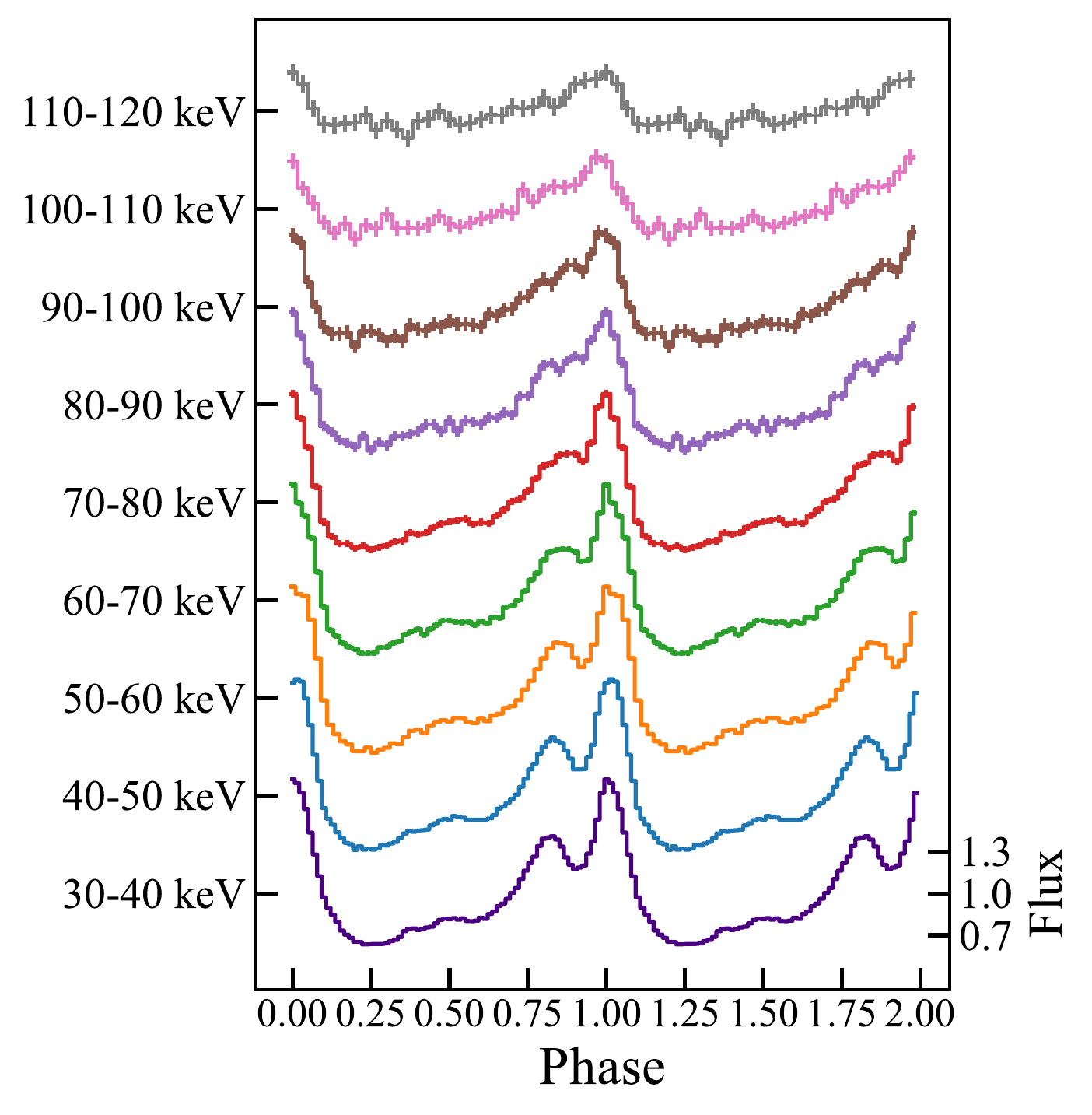}
    \includegraphics[width=1.0\columnwidth]{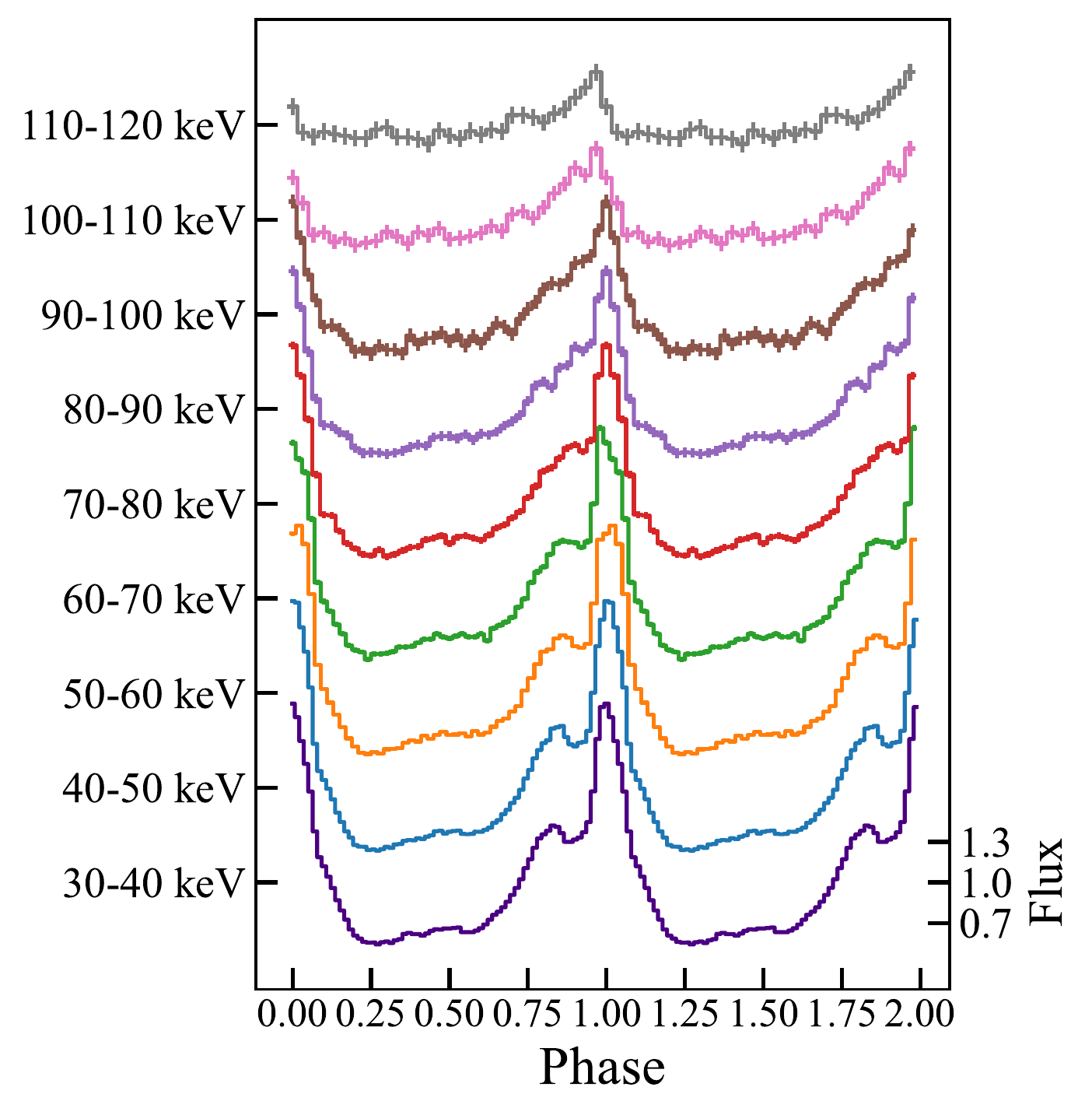}
    \caption{Same as Fig.\,\ref{fig:profiles} but using the \textit{INTEGRAL} data. 
    \textit{Left}: IntObs1. \textit{Right}: IntObs2.}
    \label{fig:integral_profiles}
\end{figure*}

The resulting evolution of the observed pulse profiles with energy is presented in Fig.~\ref{fig:profiles}.  In NuObs1 case, the pulse profiles exhibit a single-peak at low energies and a double-peak structure at high energies with the main peak at phases 0.7--1.3. A complex fine structure at the top of it and a secondary peak at phases 0.3--0.7 is also observed. As the energy increases, the relative contribution of the secondary peak grows. At energies above 6\,keV, the main peak gradually splits into two wings (phases 0.70--0.98 and 0.98--1.30). Above 22\,keV, the right wing begins to grow sharply relative to the left wing, followed by the disappearance of the latter. At phases 1.10--1.15, the previously observed dip \citep{Palombara2012, Tsygankov2012} is present, but its depth is gradually decreasing up to 26\,keV where it finally disappears, and a very pronounced right-wing peak appears in its place. The latter feature then becomes more and more prominent as the energy increases. 

\begin{figure}
    \centering
    \includegraphics[width=1.0\columnwidth]{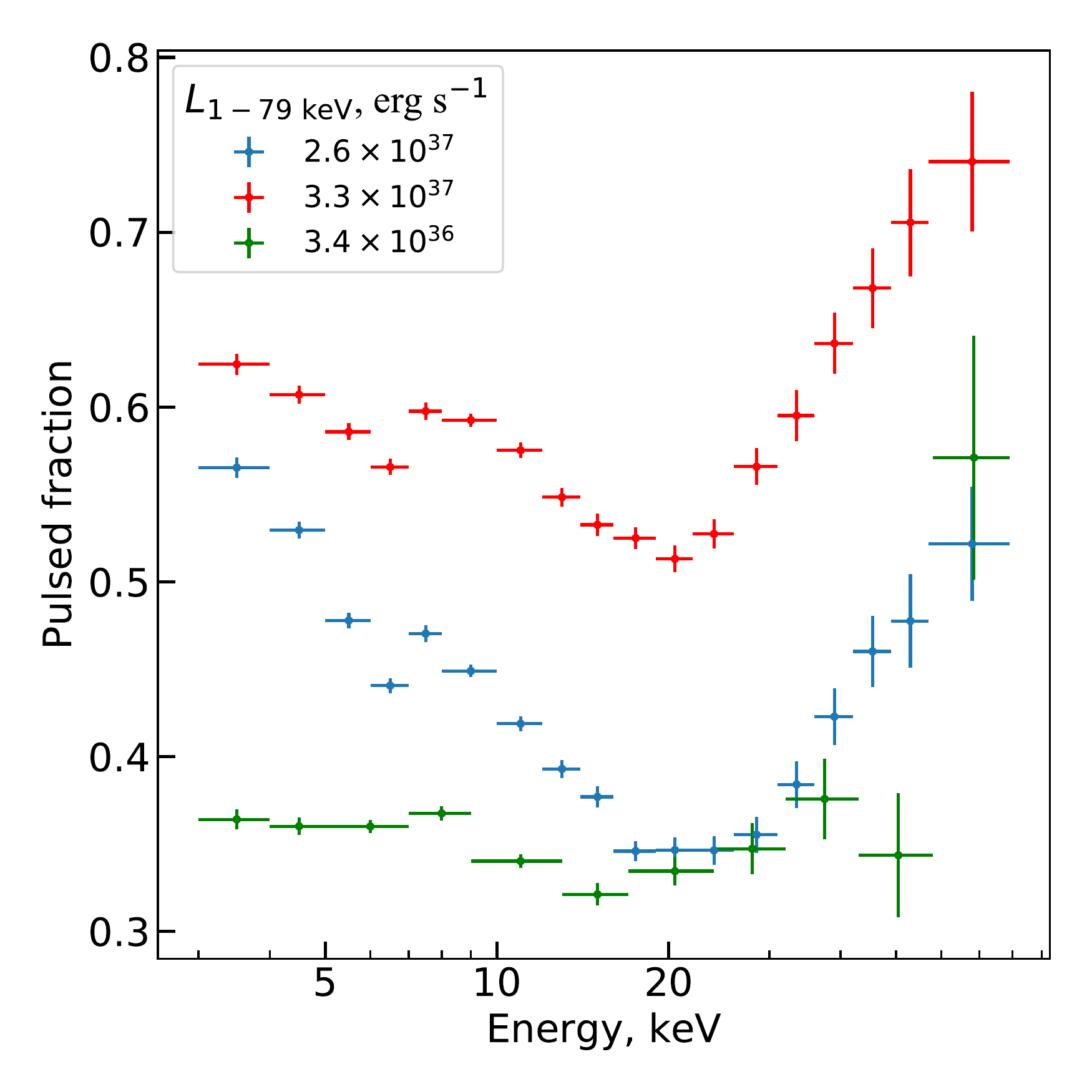}
    \caption{PF as a function of energy according to the \textit{NuSTAR} data at different luminosities.}
    \label{fig:pulsed}
\end{figure}

In NuObs2 the pulse profiles demonstrate somewhat different shape. The first thing to notice is a very well-pronounced two-wing structure of the main peak throughout the full energy range: the left one at phases 0.6--0.9 and the right one at phases 0.9--1.3. The height of the right wing increases relative to the left one  throughout 3--79\,keV energy range. The previously mentioned secondary peak around phase $\sim$0.5 does not appear in this observation neither is the dip at phase $\sim$1.1. 

The observation at significantly lower luminosity state (NuObs3) is qualitatively similar to NuObs1. The profiles show a two-peak structure with the main peak at phases 0.8--1.5 and a secondary peak at phases 0.5--0.8. At higher photon energies, the secondary peak increases relative to the main one. The previously observed dip at phase 1.2 is present in the profile at energies up to 32\,keV.

The \textit{INTEGRAL} data allowed us to study the pulse profiles in an extended range up to 120\,keV. Pulse profiles in the case of IntObs1 and IntObs2 are very similar to the corresponding profiles of NuObs2 (Fig.~\ref{fig:integral_profiles}). Both cases show a two-peak structure. The main peak consists of two wings at phases 0.6--0.9 and 0.9--1.3. The weaker secondary peak is located at phase interval 0.3--0.6.

An important metric characterising the configuration of emitting regions and radiation pattern is the dependence of the pulsed fraction (PF) on energy, usually defined as [max(rate)-min(rate)] / [max(rate) + min(rate)]. PF  calculated from the energy-resolved pulse profiles in 20 phase bins exhibits behaviour atypical for X-ray pulsars as shown in Fig.~\ref{fig:pulsed}. We note that a vast majority of bright X-ray pulsars demonstrate a monotonic increase in PF with increasing energy \citep{LutovinovTsygankov2009}.
In NuObs1 case, PF gradually drops from 56 per cent to 34 per cent in the 3--20\,keV band with a minor dip at the iron line energy (6--7\,keV range) and then rises from 34 per cent at 20\,keV to 52 per cent at 79\,keV. This is in agreement with the drop in PF in the range of 0.6--20\,keV indicated by \citet{Tsygankov2012}. In NuObs2 case, the PF demonstrates similar behaviour, first dropping from 63 per cent at 3\,keV to 52 per cent at 20\,keV and then rising to 73 per cent at 79\,keV, with the same minor dip at 6.4\,keV iron line energy. The PF in NuObs3 at low energies is significantly lower compared to the brighter states and stays  at an approximately constant level of 36 per cent below 10\,keV. At higher energies, its behaviour  resembles behaviour observed in the other two \textit{NuSTAR} observations.

\begin{figure}
    \centering
    \includegraphics[width=1.0\columnwidth]{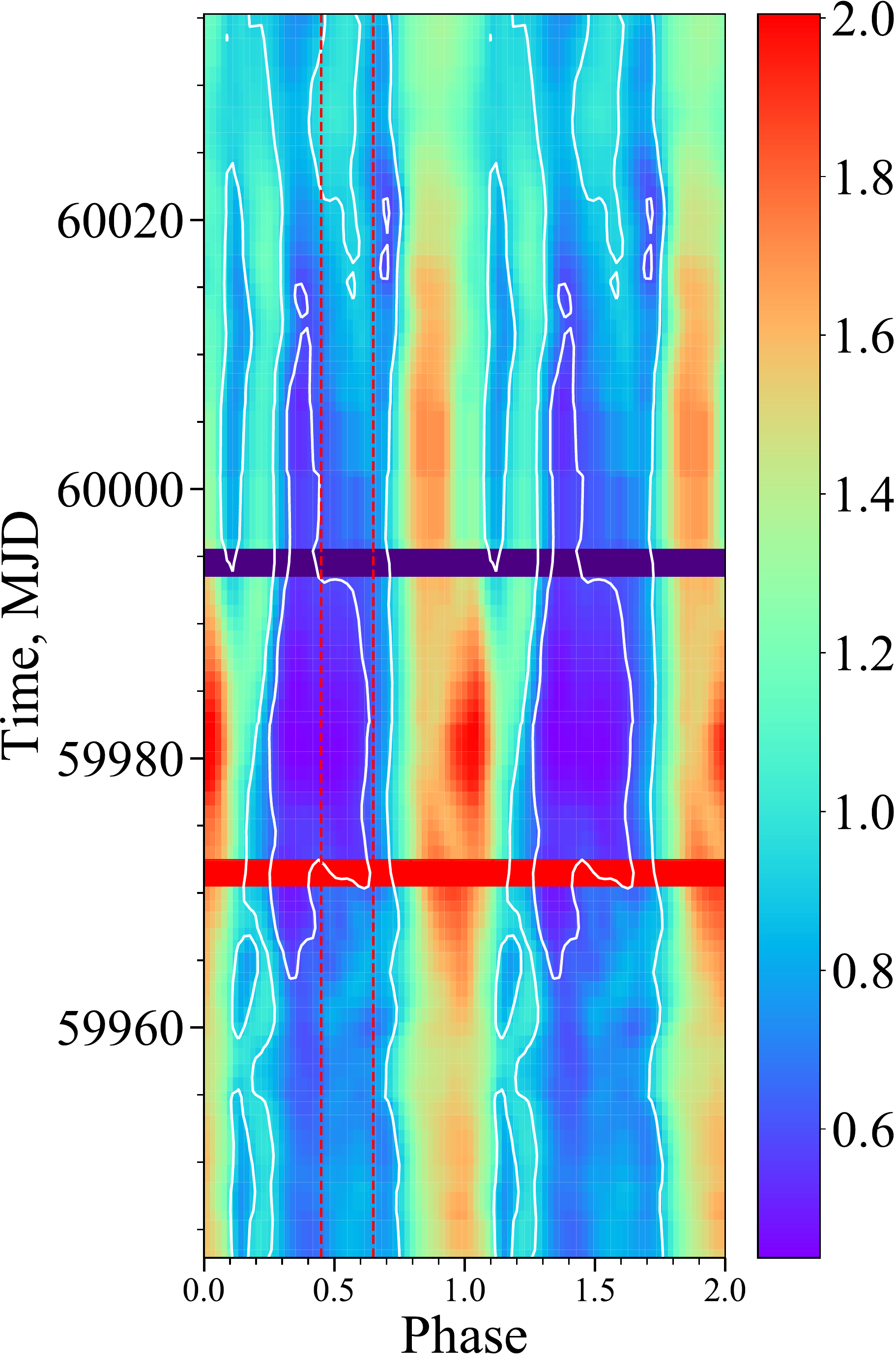}
    \caption{Two-dimensional normalized 0.3--12\,keV pulse profiles of \rx\ based on the NICER data. The vertical red dashed lines indicate approximate phase boundaries for the secondary profile peak. The horizontal red strip shows the transition from a two-peak structure to a single-peak and the indigo strip shows the backward transition. White contours are superimposed for visual convenience. The colour bar indicates values of the normalized flux in the pulse profiles. }
    \label{fig:nicer}
\end{figure}

\begin{table*}
\caption{Spectral parameters of the \texttt{comptt + comptt} model applied to the broadband observations of \rx.}
\begin{center}
\begin{tabular}{lcccccc} 
 \hline 
 & \multicolumn{6}{c}{Observation}  \\
 \cline{2-7}
  &Sep 2010 & NuObs1 & IntObs1 & IntObs2 & NuObs2 & NuObs3\\
 Parameter   && 00089583010 &  &  &  & 00089583030 \\
    \hline
 $N_{\rm H}$,~10$^{22}$ cm$^{-2}$ &&  $1.0\pm0.2$&   &   &    &  $0.24\pm0.05$ \\
 $T_{\rm bb}$, keV &  &$0.36\pm0.02$ &   &   &    &   $0.93\pm0.04$ \\
 Norm$_{\rm bb}$ $10^{-3}$\,ph\,keV$^{-1}$\,s$^{-1}$\,cm$^{-2}$ &&  $8\pm2$ &   &   &    &   $5.5^{+0.5}_{-0.4}$ \\
 $T_0$,~keV &$1.16\pm0.02$&  $1.18\pm0.01$  &
 $0.8\pm0.1$ &  $0.9\pm0.1$ &  $0.99\pm0.01$ &  $1.61^{+0.05}_{-0.04}$ \\
 $T_{\rm e, low}$,~keV &$7\pm1$& $5.21\pm0.05$   & $5.6\pm0.3$  & $5.5\pm0.3$  & $5.46\pm0.04$ &   $8.6^{+0.6}_{-0.5}$  \\ 
 $\tau_{\rm low}$  &$3.4\pm0.4$& $6.59\pm0.05$ &   $6.3^{+0.3}_{-0.4}$&  $6.4\pm0.3$&  $6.90\pm0.04$  & $3.0\pm0.1$   \\
 $T_{\rm e, high}$,~keV &$16.9^{+4.5}_{-2.2}$& $11.0\pm0.1$  &  $11.0\pm0.2$ & $11.0\pm0.2$  & $10.8\pm0.1$ &  $20.3^{+4.0}_{-2.4}$ \\ 
 $\tau_{\rm high}$  & $\geq 100$&   $\geq 100$ &  $\geq 100$ &  $\geq 100$ &  $\geq 100$  &  $\geq 100$  \\
 $E_{\rm gau}$,~keV  &&  $6.33\pm0.01$ &  & &  $6.38\pm0.01$  &  $6.31\pm0.01$  \\
 EW$_{\rm gau}$,~keV &&  $0.099\pm0.003$ &&& $0.148\pm0.004$ &  $0.040\pm0.002$ \\
 $\sigma_{\rm gau}$,~keV &&  $0.24\pm0.01$ &&& $0.38\pm0.01$ &  $0.17\pm0.03$ \\
 
 Norm$_{\rm gau}$, $10^{-3}$\,ph\,keV$^{-1}$\,s$^{-1}$\,cm$^{-2}$ &&  $10.6\pm0.3$ & & &  $20.0\pm0.6$ &  $0.97\pm0.01$  \\
 $F_{1-79\,\rm keV}$, $10^{-8}$ \flux  &$0.088\pm0.001$&  $3.61 \pm 0.01$ &  $4.4\pm0.2$ &  $3.9\pm0.2$ &
  $4.641 \pm 0.006$  &  $0.476\pm0.001$\\
 $L_{1-79\,\rm keV}$, $10^{37}$ \ergs  &$0.063\pm0.001$&  $2.573\pm0.005$ &   $3.1\pm0.2$ &  $2.8\pm0.2$ &  $3.307\pm0.004$ &  $0.339\pm0.001$ 
 \\
  $\chi^2$/d.o.f. &373/295&  3392/3237 &  47/40  &  46/40 &  3034/2784 &  3025/2765 \\
 \hline
\end{tabular}
\label{table:spec_params}
\end{center}
\end{table*}
%======================================================

\subsubsection{Pulse profile dependence on  luminosity}
To study the evolution of the pulse profile shape as a function of the luminosity during the outbursts, we constructed the time-phase matrix using the normalized pulse profiles in the 0.3--12\,keV band based on the NICER data (Fig.~\ref{fig:nicer}). We folded the NICER light curves with GAPP cubic-spline-interpolated periods following previously mentioned procedure (see Sect.~\ref{sec:broad_timing}). The phase of maximal flux was taken as the null phase for the profiles. Profiles for different observations were co-aligned using points of maximal linear cross-correlation as references. The matrix shows the transition of soft-energy pulse profiles from the double-peaked structure to the single-peaked structure at approximately MJD 59971 (see  Fig.~\ref{fig:nicer_profiles}). Subsequently a reverse transition is observed around MJD 59995. The previously mentioned dip in the NICER and \textit{NuSTAR} profiles disappears when 1--10\,keV luminosity $L_{\rm 1-10}$ exceeds $5.9\times10^{36}$ \ergs\ (MJD 59969) (see Fig.~\ref{fig:profiles} and closed white contours at phases 0.0--0.2, 1.0--1.2 in Fig.~\ref{fig:nicer}). The dip reappears at a luminosity below $7.6\times10^{36}$ \ergs\ (MJD 59995), which is close to the luminosity at which it disappeared during the rising part of the outburst. 

\begin{figure}
    \centering
    \includegraphics[width=0.7\columnwidth]{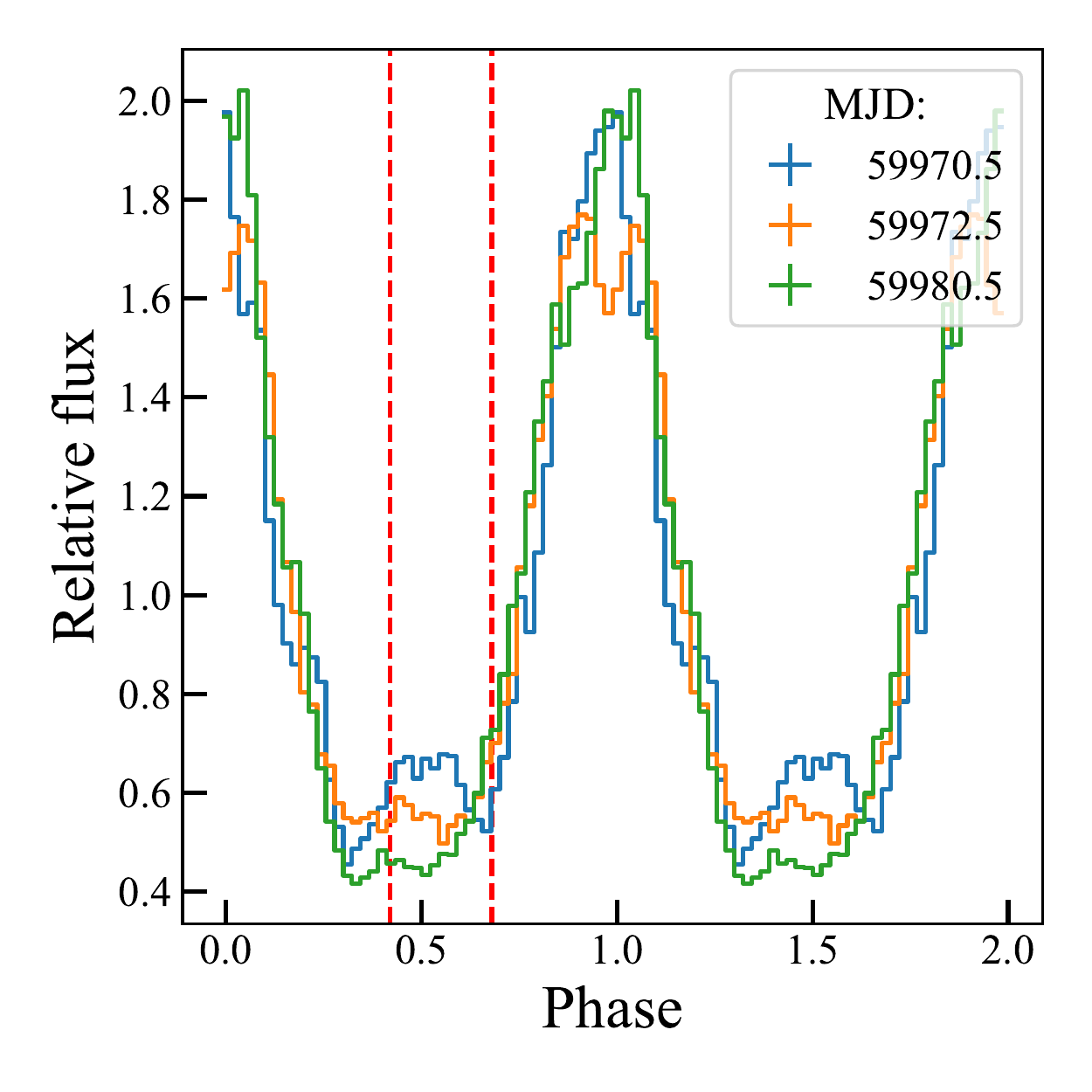} 
    \caption{Normalized pulse profiles of \rx\ in the 0.3--12\,keV band during the transition from a two-peak structure to a single peak based on the NICER data. The red vertical dashed lines indicate the approximate phase boundaries for the secondary peak. }
    \label{fig:nicer_profiles}
\end{figure}

\begin{figure}
    \centering
    \includegraphics[width=1.0\columnwidth]{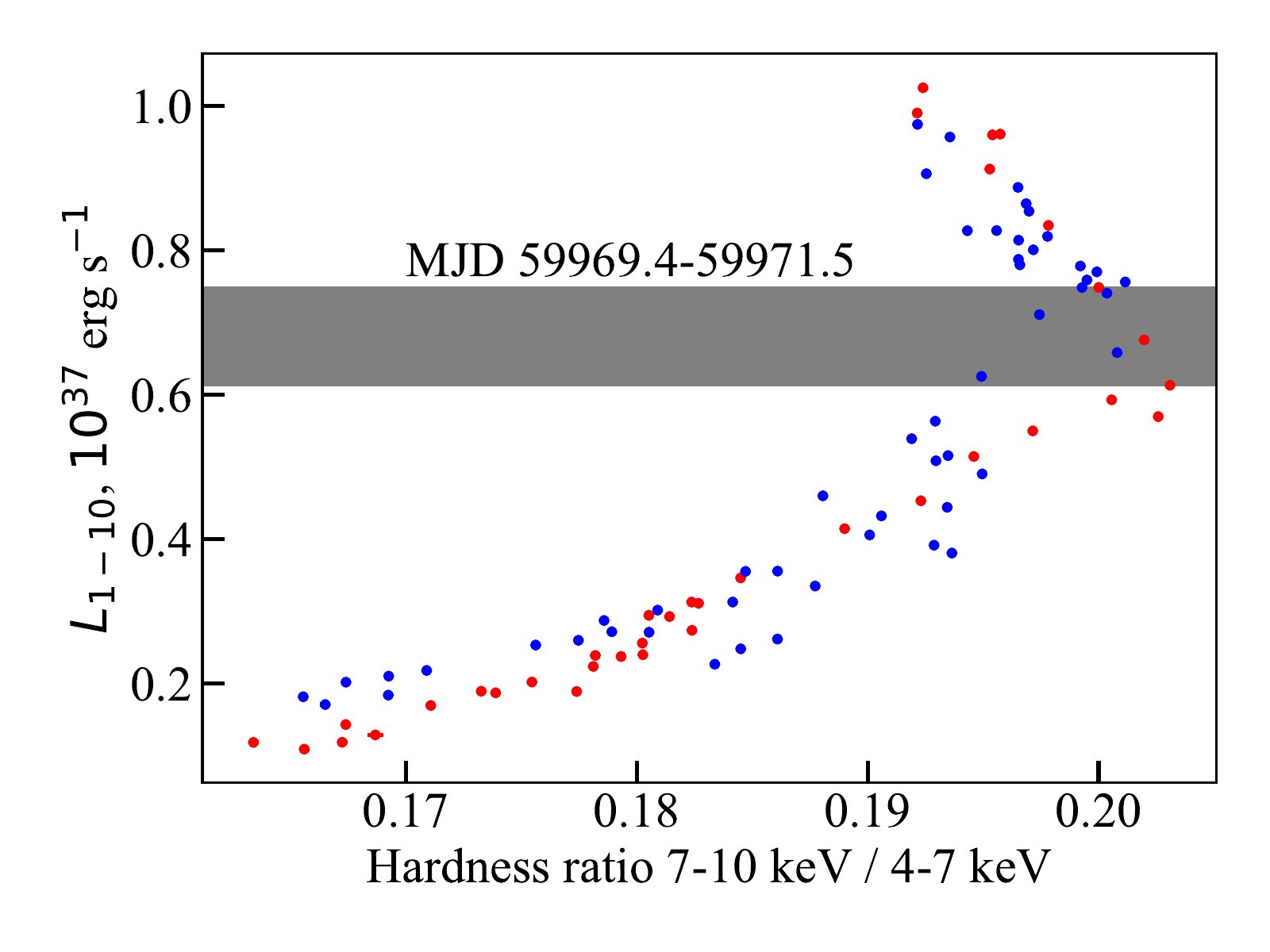}\\
    \caption{Dependence of the hardness ratio (7--10\,keV/4--7\,keV) on the 1--10\,keV luminosity based on the NICER data. The grey strip indicates the transition between branches.  The red points correspond to the outburst rise, while the blue points are for the outburst decay. }
    \label{fig:hardness}
\end{figure} 
%======================================================

We then investigated the dependence of the 7--10\,keV/4--7\,keV hardness ratio on the luminosity using the NICER data in order to study further the spectral transition of \rx\ that occurred around MJD 59971 \citep{Coley2023}. It was shown previously that such a dependence can be used to deduce the regime of accretion in the X-ray pulsars \citep[see][]{Reig2013}. The resulting pattern (see Fig.~\ref{fig:hardness}) consists of two branches.  
Here to convert the NICER 1--10\,keV luminosity $L_{\rm 1-10}$ to the bolometric one $L_{\rm X}$ (1--79\,keV), we used a bolometric correction of 4.15, the ratio of the NICER and \textit{NuSTAR} fluxes during NuObs2 observation. Similarly, we calculated the bolometric peak luminosity during the outburst $L_{\rm X}=4.3\times10^{37}$\ergs.
The transition between branches occurred first at MJD 59971 at the luminosity $L_{\rm X}\sim2.8\times10^{37}$\ergs\ during the brightening phase of the outburst. 
The opposite transition occurred on MJD 59995 at an approximately the same luminosity during the decay of the outburst. 
This is in a good agreement with the times of transition from a double-peak to a single-peak structure of the 0.3--12\,keV pulse profile and a subsequent transition back to a one-peak structure during the rising and decay phases, respectively. We argue thus that both transitions have the same physical origin and are likely associated with a transition of the pulsar to the supercritical accretion state.

\begin{figure}
    \centering
 	\includegraphics[width=1.0\columnwidth]{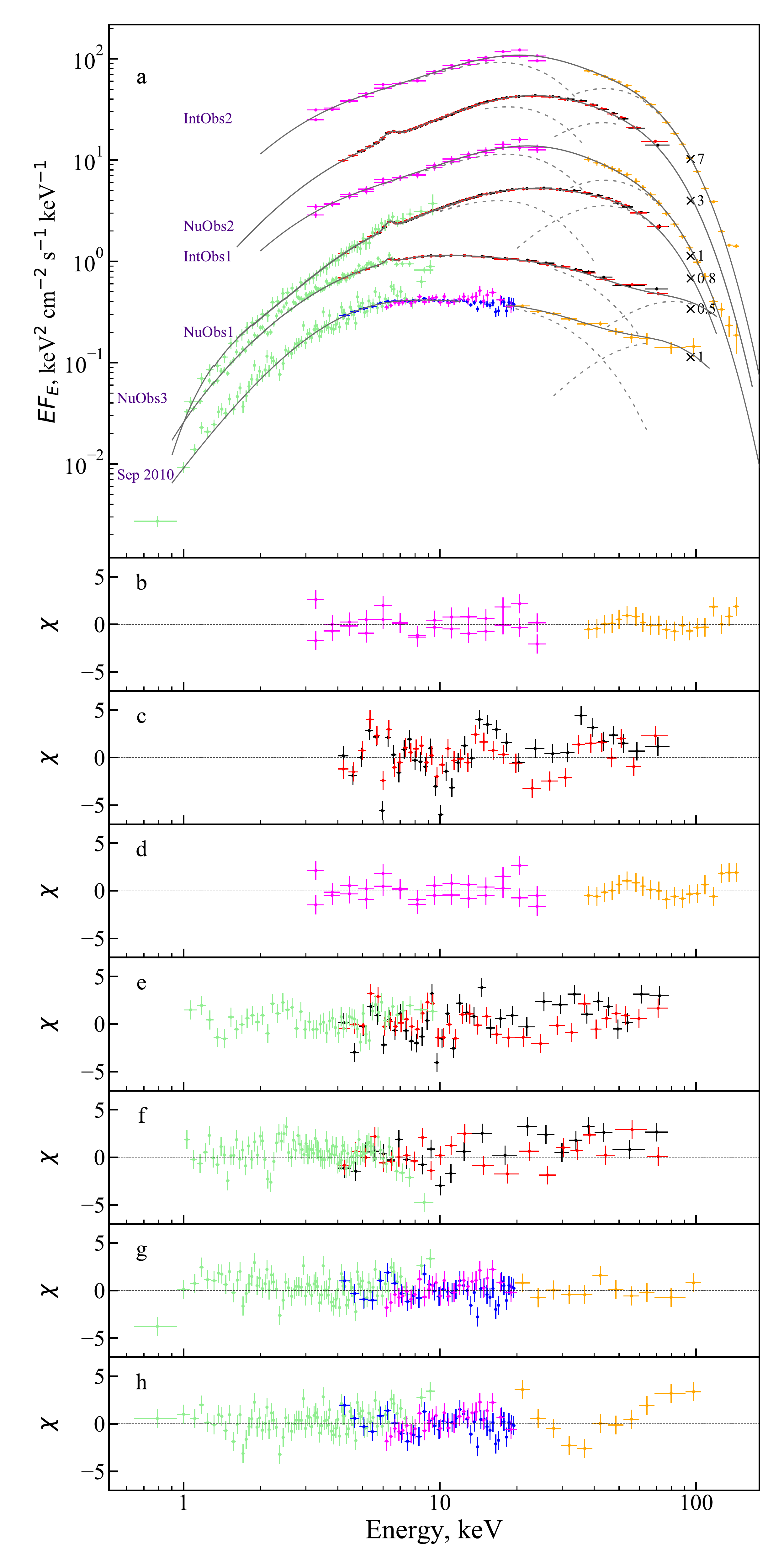}
 	\caption{Unfolded spectrum of \rx\ during outbursts in January-April 2023 and in September 2010 \citep{Tsygankov2012} and its approximation with the two-hump \texttt{comptt+comptt} continuum model (solid lines in panel a), dashed lines represent humps separately. Red and black crosses are for the FPMA and FPMB telescopes respectively, light green for the {\it Swift}/XRT telescope, magenta crosses are for JEM-X2, blue for the {\it RXTE}/PCA, and orange for the {\it INTEGRAL}/IBIS. 
  The spectra were spaced along the y-axis with the factors shown for visual convenience.
    The bottom panels show the residuals for different observations and continuum models: two-hump (IntObs2, panel b), two-hump (NuObs2, panel c), two-hump (IntObs1, panel d), two-hump (NuObs1, panel e), two-hump (NuObs3, panel f),
    two-hump (September 2010, panel g), \texttt{cutoffpl + bbodyrad} (September 2010, panel h).  }
	\label{fig:spec}
\end{figure}

\subsection{Spectral analysis}
\label{sec:spectrum}
During the Type II outburst of \rx\ in 2023, five broadband observations by \textit{NuSTAR} and \textit{INTEGRAL} were made near the peak of the outburst.  We used the 4--79\,keV \textit{NuSTAR} data for the spectral analysis. For the \textit{Swift} data, the 1--10\,keV range was used. Data from \textit{INTEGRAL}/ISGRI (35--150\,keV), \textit{INTEGRAL}/JEM-X1, and \textit{INTEGRAL}/JEM-X2 (3--30\,keV) were also used for analysis.
To expand the energy range covered, we jointly modelled the \textit{NuSTAR} observation NuObs1 with its nearby \textit{Swift}/XRT observation (ObsID 00089583010, MJD 59969), as well as NuObs3 with its nearby \textit{Swift}/XRT observation (ObsID 00089583030, MJD 60041). NICER/XTI data were not used for joint approximation due to its cross-calibration difficulties with the \textit{NuSTAR} and \textit{INTEGRAL} data. To take into account inaccuracies in the calibration between the detectors, we introduced the cross-calibration multiplicative component \texttt{const} (all model names correspond to those in {\sc XSPEC} package).

To model broadband continuum spectrum we first considered several single-component models such as power law with an exponential cutoff (\texttt{cutoffpl}), as well as the Comptonization model \texttt{comptt} \citep{Titarchuk1994}. To account for the interstellar absorption, we also included a multiplicative component \texttt{tbabs} with the abundances from \citet{Wilms2000}. A strong K$\alpha$ iron line which is present regardless of the choice of the continuum model was modelled using an additive \texttt{gauss} component. None of the one-component continuum models gave a satisfactory description of the spectrum and we added a 30~keV cyclotron line following \citet{Tsygankov2012}, which led to some improvement in the approximation accuracy. 
However, still none of these models resulted in an acceptable fit, with the reduced chi-squared $\chi^2_{\rm red}\gtrsim1.2-1.3$. 
We thus decided to consider two-component continuum models \texttt{comptt + comptt} \citep[see][]{Tsygankov2019b} and \texttt{cutoffpl+bbodyrad}. The latter one was used by \citet{Tsygankov2012} to describe \rx\ spectrum in the low-luminosity state (see Fig.~\ref{fig:spec}h). Of the two two-component continuum models, the best fit is provided by the \texttt{comptt+comptt} model for all broad-band observations (see Table~\ref{table:spec_params}). An additional argument against choosing the \texttt{cutoffpl+bbodyrad} model is a high blackbody temperature of $T=(3.7\pm0.1)$~keV, which is not typical for Be/X-ray systems \citep{Hickox2004}. In turn, the peaks of the \texttt{comptt+comptt} model at $\sim$20\,keV and $\sim$45\,keV agree with the typical hump positions in other X-ray pulsars \citep[see, e.g.,][]{Doroshenko2012, Tsygankov2019a,Tsygankov2019b, Lutovinov2021, Doroshenko2021, Doroshenko2022}.
An extra low-temperature blackbody component \texttt{bbody} was added to the continuum in observations NuObs1 and NuObs3. The results of approximation of \textit{INTEGRAL} and \textit{NuSTAR} spectral data are shown in Fig.~\ref{fig:spec}. 
In the case of NuObs3 and September 2010 \textit{INTEGRAL} observations at luminosities of $\sim10^{36}$ \ergs\ the two-hump model components peak at energies of $\sim$10 and $\sim$70\,keV. The seed photon temperatures $T_0$ of both humps were tied together. Within the framework of this model, the optical depth of the high-energy hump cannot be constrained (i.e. has value $>100$), which makes its use in the model almost equivalent to using a blackbody. However, to keep the consistency with the previously published works, we  continue to follow the generally accepted practice of describing such spectra with \texttt{comptt+comptt} model.

A large optical depth of the high-energy component implied by the fit is likely related to the nature of this component, which is produced by multiple Compton scatterings of the seed cyclotron photons in the neutron star atmosphere  \citep{Mushtukov2021, SokolovaLapa2021}.
The typical emission depth of seed cyclotron photons is related to the braking distance of the accretion flow in the neutron star atmosphere. 
Because Compton scattering is resonant around the cyclotron energy, the scattering cross section can be orders of magnitude above the Thomson value \citep{1986ApJ...309..362D, 1991ApJ...374..687H}. 
Under this condition, the photon mean free path is very small, and each photon undergoes many scattering before it is finally able to leave the stellar atmosphere.
Resonant scattering of photons by hot electrons allows photons to experience significant changes in their energy within the Doppler core of the cyclotron line.
As a result, many photons leave the atmosphere in the wings of the line where the cross-section is smaller and chances to leave the atmosphere are larger which leads to the formation of a broad high-energy component. 
However, even for photons leaving the atmosphere in the wings, the typical number of scatterings remains large because in each scattering photons tend to get energy close to the cyclotron one, i.e. photons are confined near the cyclotron energy \citep{2022PhRvD.105j3027M}. 
%In contrast, the probability of scattering into the wings is small.

The results of the spectral approximation by the \texttt{comptt+comptt} continuum model are given in Table \ref{table:spec_params}.  The addition of the \texttt{tbabs} component improved the quality of the approximation only in the case of NuObs1 and NuObs3, where it was therefore taken into account. The obtained hydrogen column density $N_{\rm H}$ turned out to be roughly consistent with the Galactic value in the direction to the source \citep[$0.6\times10^{22}$\,cm$^{-2}$; ][]{HI4PI2016}.
Regardless of the two-component continuum model, the presence of a cyclotron resonant scattering feature (CRSF) was not required in the energy range of 1--79\,keV. 

The high quality of the \textit{NuSTAR} data allowed us to study the evolution of the spectral parameters with the pulse phase in NuObs1 and NuObs2. We divided the \textit{NuSTAR} data into 10 evenly distributed phases. The obtained phase-resolved spectra were approximated by the same model as phase-averaged spectra (\texttt{const$\times$(comptt+comptt+gauss)}, see Fig.~\ref{fig:phres_spec}).  
The optical depth of the high-energy hump was fixed, however, at $\tau=200$ for phase-resolved analysis. Both NuObs1 and NuObs2 exhibit similar patterns in the evolution of their spectral parameters. The flux shows a strong correlation with the seed photon temperature $T_0$ and an anti-correlation with the low-energy hump temperature $T_{\rm low}$. The high-energy hump temperature $T_{\rm high}$ dependency on the phase demonstrates a two-peak structure with the peak maxima at phases 0.55 and 1.05. The energy of the iron line $E_{\rm gau}$ did not show significant changes depending on the phase in the case of NuObs1, while in the case of NuObs2 there is a strong correlation with flux in the profile. The iron line normalization Norm$_{\rm gau}$, the equivalent width EW$_{\rm gau}$, and the iron line width $\sigma_{\rm gau}$ all have well-pronounced sinusoidal shape and demonstrate strong anti-correlation with the flux.  Possible phase-transient cyclotron lines \citep[see, e.g.,][]{Kreykenbohm2002, Doroshenko2017, Molkov2019, Molkov2021, Salganik2022} were not found in any of the spectra. 

To assess changes of spectral hardness with pulse phase we examined a ratio of the unnormalized count rates in the 10--20\,keV to that in the 3--10\,keV range and 20--30\,keV to 10--20\,keV (Fig.~\ref{fig:phres_spec}). 
As evident from the figure, there is a strong anti-correlation between hardness ratios and the flux with the linear cross-correlation coefficient of $-0.74$ for 10--20\,keV/3--10\,keV and $-0.72$ for 20--30\,keV/10--20\,keV for NuObs1. We note that this pattern is opposite to that observed in the case of another long-period pulsar eRASSU J050810.4$-$660653 \citep{Salganik2022c}, where a strong correlation  between  hardness  and  flux was observed at a similar luminosity $L_{\rm X}\sim2\times10^{37}$\,\ergs.

\begin{figure*}
    \centering
 	\includegraphics[width=0.8\columnwidth]{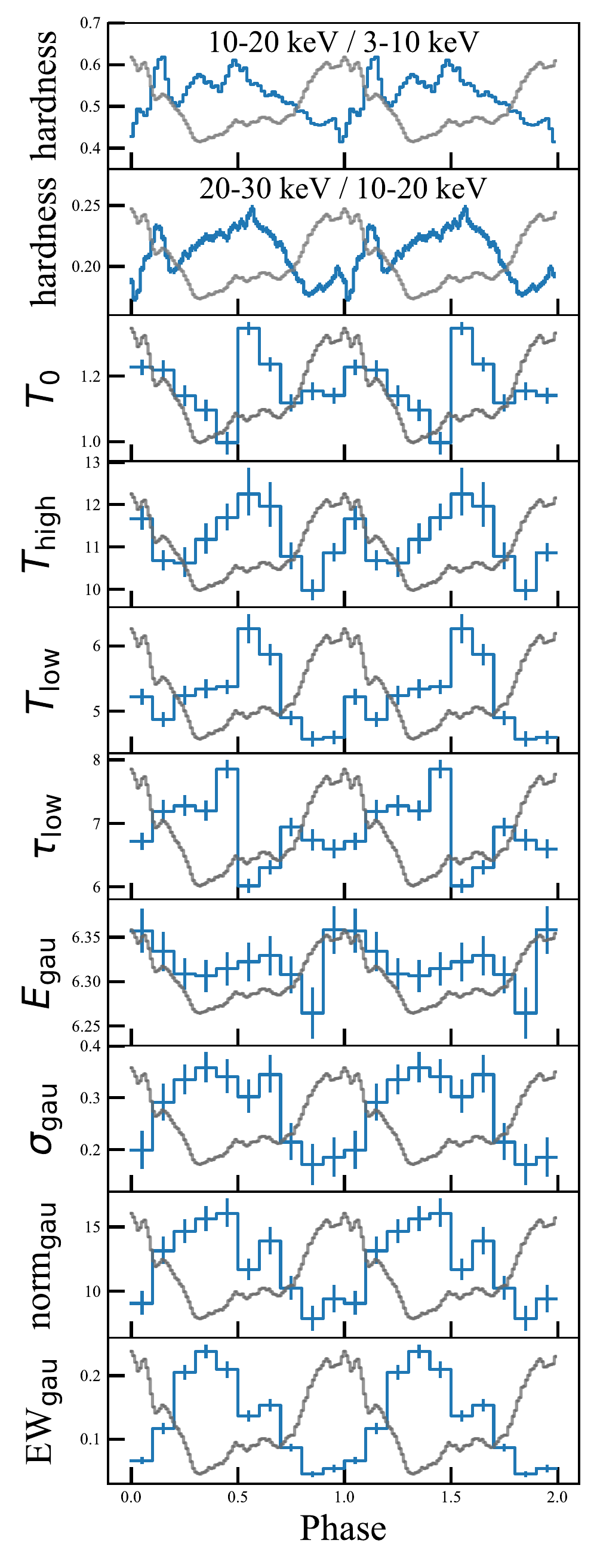} \hspace{1cm}
\includegraphics[width=0.8\columnwidth]{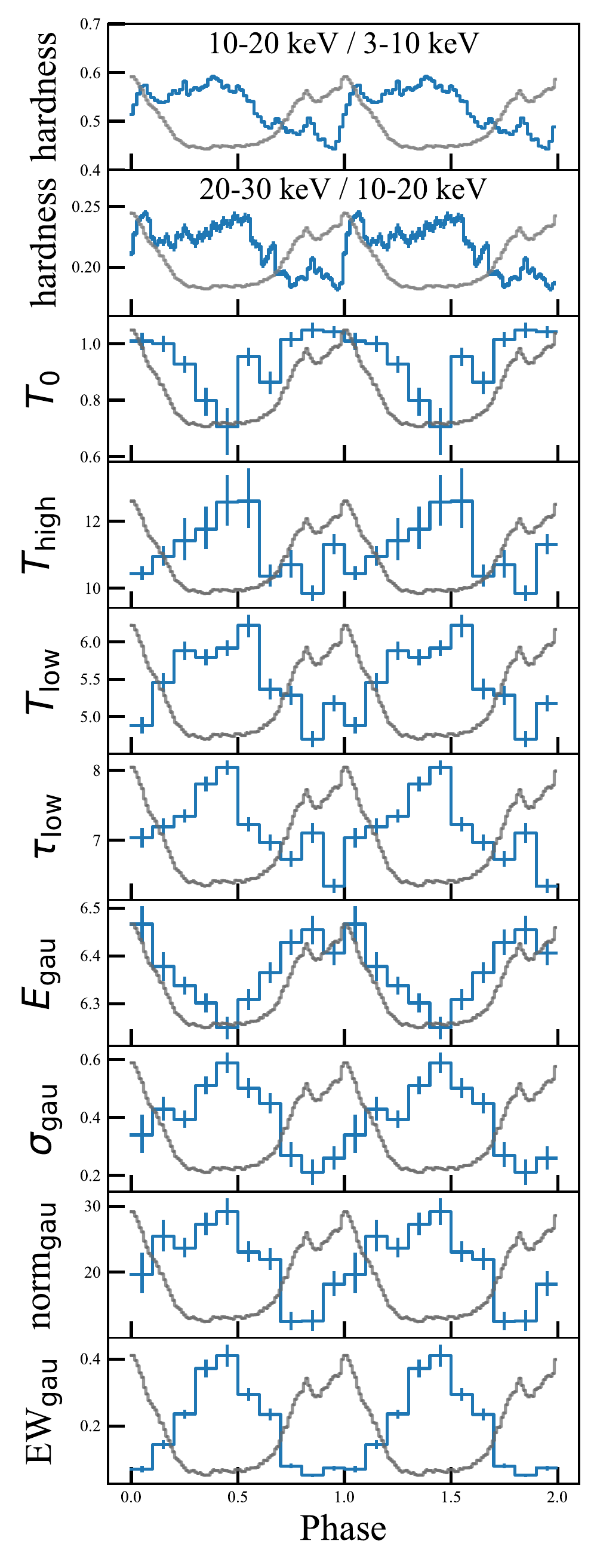}
\caption{Dependence of the hardness ratios and the spectral parameters on the pulse phase for NuObs1 \textit{(left panel)} and NuObs2 \textit{(right panel)}. 
The averaged pulse profile in a wide energy range 3--79\,keV is superimposed in grey for visual comparison. Units of the parameters values coincide with those in Table~\ref{table:spec_params}.}
  \label{fig:phres_spec}
\end{figure*}
%======================================================

\section{Discussion}

\subsection{Revision of the cyclotron line identification}
\label{sec:cycl}
As discussed above, modelling of the \rx\ spectrum in broad energy range reveals no evidence for presence of narrow absorption features which could be interpreted as a cyclotron line as earlier reported in the literature \citep{Tsygankov2012} despite significantly better counting statistics of the dataset considered here. We note that a limited sensitivity of X-ray instruments capable of broadband spectroscopy implied that a detection of cyclotron line in most cases was only possible for relatively bright pulsars with luminosities $\gtrsim10^{36}$\,\ergs. At this luminosities the choice of a power law with an exponential cutoff as the main continuum model \citep[see, e.g.,][]{Coburn2002}, which is characteristic of high accretion rates \citep{Becker2007, Farinelli2016}, is indeed feasible. However, in recent years it has been shown that as soon as the luminosity of a Be/X-ray system is in the range $10^{34}-10^{36}$ \ergs, the observed broadband continuum changes and exhibits two broad peaks in the soft ($\leq10$\,keV) and hard ($\geq20$\,keV) bands, see, e.g., \citet{Tsygankov2019a} for the case of GX 304$-$1, \citet{Tsygankov2019b} for A~0535+262  and \citet{Lutovinov2021} for GRO~J1008$-$57. Here each of the components can be roughly approximated by the Comptonization model from \citet{Titarchuk1994}. We note that a similar spectral shape is observed at somewhat higher luminosities, albeit in a much less pronounced form \citep{Tsygankov2019b, Doroshenko2020}.  As a result of such spectral transitions at low luminosities, a deficit of photons naturally arises between these two spectral components and can be interpreted as an absorption feature, especially, if counting statistics is modest. 

\citet{Tsygankov2012} used a combination of a cutoff power law  (\texttt{cutoffpl}) and a blackbody  (\texttt{bbodyrad}) for the approximation of the continuum of \rx\ in the low-luminosity state. This resulted in a deficit of photons around 30\,keV in the spectrum  relative to the model, which was interpreted as the presence of CRSF (Fig.~\ref{fig:spec}h). However, when the continuum can in fact be approximated by the previously mentioned \texttt{comptt}+\texttt{comptt}  spectral model with the peaks at 10 and 70\,keV, there is no deficit of photons (Fig.~\ref{fig:spec}g). 
As we showed in Sect.~\ref{sec:spectrum}, the continuum spectrum of \rx\ in the bright state is well described by the two-component models \texttt{bbodyrad+cutoffpl} and \texttt{comptt+comptt} not showing any photon deficit around 30\,keV. A similar situation was observed in the case of a pulsar KS~1947+300  \citep{Doroshenko2020}, when a single spurious CRSF arose due to a continuum fit by a combination of the \texttt{cutoffpl} and \texttt{bbodyrad} models, but disappeared when a two-component \texttt{comptt+comptt} model was used. The absence of this deficit at high luminosities suggests that the gap between the humps observed at $\sim$20 times lower luminosity \citep{Tsygankov2012} and previously interpreted as the presence of CRSF is in fact of non-magnetic nature.

\subsection{Indirect magnetic field estimates}

Considering the conflicting reports regarding presence of a cyclotron line in the spectrum of the source, it is worth to estimate the magnetic field strength using other arguments some of which are discussed below.

\subsubsection{Spin-up and accretion torque}

The accretion torque acting upon the neutron star is proportional to the accretion rate and size of the magnetosphere. Thus the magnetic field can be estimated based on the observed maximum spin-up rate close to the peak of the giant outburst where any braking torques can be ignored. Although no orbital solution exists as of moment of writing, it is reasonable to assume that the spin evolution at this point is driven predominantly by accretion torque, which is confirmed by a strong correlation between the observed spin frequency derivative and flux which can be constructed as described in \citet{2016A&A...589A..72D} and shown in Fig.~\ref{fig:spinup}. 
The maximum observed spin-up  is $\dot{\nu}\sim2.8\times10^{-11}$\,Hz\,s$^{-1}$ at a bolometric luminosity of $\sim4.1\times10^{37}$\,\ergs\ or accretion rate of $\sim2.6\times10^{17}$\,g\,s$^{-1}$ assuming a standard neutron star parameters, mass of $M=1.4$M$_{\sun}$ and radius of $R=12$ km. The accretion-induced spin-up (neglecting spin-down torques) is defined by the inner radius of the accretion disc and accretion rate, i.e. $I\dot{\Omega}=\dot{M}\sqrt{GMR_{\rm d}}$ or $R_{\rm d}=(I 2\pi\dot{\nu}/\dot{M})^2/GM\sim2\times10^{9}$\,cm for the moment of inertia of $I=10^{45}$\,g\,cm$^2$. 
The lower limit on the magnetic field strength can then be derived assuming that the disc is truncated at magnetosphere, i.e. the disc radius equals Alf\'ven radius $R_{\rm d}=R_{\rm A}=[\mu^2/(2\dot{M}\sqrt{2GM})]^{2/7}$, which yields $\mu\approx 6\times10^{31}$\,G\,cm$^{3}$ or $B\approx 3.5\times10^{13}$\,G. Note that in reality the disc likely pushes deeper into the magnetosphere, i.e. $R_{\rm d}=kR_{\rm A}$ with $k\sim0.2-0.5$ \citep{GhoshLamb1979b,2020A&A...643A..62D}, which correspondingly would require even stronger field of up to $\sim10^{14}$\,G. 

\subsubsection{Accretion regime transition}
The observed luminosity of the accretion regime transition is defined by the local accretion rate and thus by the magnetic field strength \citep{1976MNRAS.175..395B} and thus can also be used to estimate the field.  We measure critical luminosity of $L_{\rm c}\sim2.8\times10^{37}$\,erg~s$^{-1}$ from the observed turn-over point in the hardness-luminosity diagram (Fig.~\ref{fig:hardness}) as well as from the observed transition from a two-peak to a single-peak structure of the low-energy pulse profiles (Figs~\ref{fig:nicer} and \ref{fig:nicer_profiles}).

From a theoretical perspective, \citet{2012A&A...544A.123B} estimated $L_{\rm c}\sim1.5\times 10^{37}\,(B/10^{12})^{16/15}$\,erg~s$^{-1}$.
Therefore, the field of the order of $10^{12}$\,G is fully consistent with the observed critical luminosity. We need to keep in mind, however, that there is a considerable uncertainty in theoretically predicted value of the transition (aka critical) luminosity. 
For instance, critical luminosity estimates by \cite{2015MNRAS.447.1847M} imply field of the order of $10^{13-14}$\,G (depending on assumed geometry), which appears to be more in line with the estimate based on maximal observed spin-up rate quoted above. 

\subsubsection{Cold disc accretion luminosity}

It is interesting that the source flux in quiescence has been consistently reported to be relatively high \citep{Reig1999, Palombara2012}. This is not surprising given the long period of the source, which implies that upon transition to quiescence the accretion is not expected to be centrifugally inhibited, but rather a transition to a meta-stable accretion state when the pulsar accretes from a cold non-ionized disc can be anticipated \citep{Tsygankov2017_2}. The accretion luminosity is expected in this case to be:
\begin{equation}
L\le L_{\rm cold}=9\times10^{33}k^{1.5}M_{1.4}^{0.28}R_6^{1.57}B_{12}^{0.86}\ \mathrm{erg\,s}^{-1}, 
\end{equation}
where $M_{1.4}$, $R_{6}$ and $B_{12}$ are neutron star mass, radius and magnetic field normalized to 1.4M$_\odot$, $10^{6}$\,cm and $10^{12}$\,G, respectively.
\citet{Reig1999} estimated the quiescence luminosity  of $\sim1.7\times10^{34}$\,\ergs\ (recalculated for the distance of 2.44\,kpc). 
From the equation above one can estimate that field of $B_{12}\simeq5.0$ is required to match the observed luminosity for $k=0.5$. One needs to bear in mind, however, that the luminosity reported by \citet{Reig1999} corresponds to the average value following a long period of quiescence, and that is expected to be below the transitional luminosity $L_{\rm cold}$. Indeed, \citet{Palombara2012} report somewhat higher luminosity of $\sim4.4\times10^{34}$\,\ergs\ that corresponds to $B_{12}\sim15$ if the distance of 2.44\,kpc is assumed. Again, this is in line with the other estimates quoted above and indicates a relatively strong field in excess of $10^{13}$\,G. 

\section{Summary} 
We presented here the results of the first detailed study of the spectral and temporal properties of the Be/X-ray system \rx\ during a giant outburst observed from the source in the beginning of 2023 using the data obtained by the \textit{NuSTAR}, \textit{INTEGRAL}, \textit{Swift} and NICER observatories.
The evolution of the pulse profiles and the flux hardness during the outburst showed a transition, which can be interpreted as a transition from sub- to the supercritical accretion regime associated with onset of an accretion column at the luminosity $L_{\rm c}\sim2.8\times10^{37}$\ergs.  The pulse profiles show a two-peak structure before the transition to the supercritical accretion regime, while after the transition they show a single-peak structure. The profiles show a dip at luminosities up to $L_{\rm X}\sim2.6\times10^{37}$\ergs\ which transits into a peak at energies above 26\,keV according to the energy-resolved study. The PF showed a non-monotonic character with a gradual drop to 20\,keV and a sharp increase above that energy and a minor drop at the energy of the iron line. The analysis of the pulse profile shapes and PF extends up to 120\,keV near the peak of the outburst where statistics is sufficient.

A study of broadband spectra near the outburst peak showed that the spectral continuum is best described by the two-component model \texttt{comptt+comptt} with components peaking at $\sim$10--20 and 50--70\,keV respectively depending on the source luminosity. The results of the analysis of the phase-resolved spectra showed that the temperature of soft photons $T_0$ correlates with the flux, while the temperatures of the humps $T_{\rm low}$ and $T_{\rm high}$, as well as the iron line flux and equivalent width anticorrelate with the X-ray flux. The hardness ratios 10-20/3-10\,keV and 20-30\,keV/10-20\,keV both anti-correlate with the flux. Spectral modelling of both phase-averaged and phase-resolved data do not require inclusion of the 30\,keV cyclotron line to the continuum model. This allows us to conclude that the previously observed deficit of photons around 30\,keV is of a nonmagnetic nature. The likely spurious origin of the previously reported CRSF necessitates alternative magnetic field estimates which we also discussed. In particular, we conclude that the field must be greater than $10^{13}$\,G based on the pulsar spin-up rates, the transition of the pulsar to the supercritical accretion regime, and its cold disc accretion luminosity. We note that in this case one actually does not expect a detection of the CRSF within energy band of \textit{NuSTAR} and thus non-detection of a line is not surprising.

%%%%%%%%%%%%%%%%%%%%%%%%%%%%%%%%%%%%%%%%%%%%%%%%%%%%%%%%%%%%%%%%%%%%%%%%%%%%%%
%% Acknowledgments                                                         %%
%%%%%%%%%%%%%%%%%%%%%%%%%%%%%%%%%%%%%%%%%%%%%%%%%%%%%%%%%%%%%%%%%%%%%%%%%%%%%%
\section*{Acknowledgements}

We thank Ilya Mereminskiy, Andrei Semena, and Andrey Shtykovsky for helpful discussions, and Carlo Ferrigno for the help with the {\it INTEGRAL} data reduction.
We are grateful to the \textit{NuSTAR} team for approving and rapid scheduling of the follow-up observation. We are grateful to the \textit{Swift} team for approving and rapid scheduling of the monitoring campaign. This work made use of data supplied by the UK \textit{Swift} Science Data Centre at the University of Leicester and data obtained with \textit{NuSTAR} mission, a project led by Caltech, funded by NASA and managed by JPL. This research also has made use of the \textit{NuSTAR} Data Analysis Software ({\sc NUSTARDAS}) jointly developed by the ASI Science Data Center (ASDC, Italy) and Caltech. This research has made use of data and software provided by the High Energy Astrophysics Science Archive Research Center (HEASARC), which is a service of the Astrophysics Science Division at NASA/GSFC and the High Energy Astrophysics Division of the Smithsonian Astrophysical Observatory. 
We acknowledge support from the Russian Science Foundation grant 19-12-00423 (AS, SST, SVM, AAL), the Academy of Finland grants 333112, 349144, and 349906 (JP), the German Academic Exchange Service (DAAD) travel grant 57525212 (VD), and the UKRI Stephen Hawking fellowship (AAM).

\section{Data availability}
\textit{NuSTAR}, NICER, \textit{INTEGRAL} and \textit{Swift} data can be accessed from corresponding online archives. 
%%%%%%%%%%%%%%%%%%%%%%%%%%%%%%%%%%%%%%%%%%%%%%%%%%%%%%%%%%%%%%%%%%%%%%%%%%%%%%
%% Bibliography                                                             %%
%%%%%%%%%%%%%%%%%%%%%%%%%%%%%%%%%%%%%%%%%%%%%%%%%%%%%%%%%%%%%%%%%%%%%%%%%%%%%%
 
\bibliographystyle{mnras}
\bibliography{allbib}

 % Don't change these lines
\bsp    % typesetting comment
\label{lastpage}
\end{document}